\documentclass[a4paper]{article}

\usepackage{INTERSPEECH2020}
\usepackage{todonotes}
\usepackage{enumitem}
\usepackage{multirow}
\usepackage{array}
\usepackage{balance}
\usepackage{url}

\title{Speaker Anonymization with Distribution-Preserving X-Vector Generation for the VoicePrivacy Challenge 2020}
\name{Henry Turner, Giulio Lovisotto, Ivan Martinovic}
\address{
  University of Oxford, UK}
\email{firstname.lastname@cs.ox.ac.uk}

\begin{document}

\maketitle
\begin{abstract}
In this paper, we present a Distribution-Preserving Voice Anonymization technique, as our submission to the VoicePrivacy  Challenge 2020.
We observe that the challenge baseline system generates fake X-vectors which are very similar to each other, significantly more so than those extracted from organic speakers.
This difference arises from averaging many X-vectors from a pool of speakers in the anonymization process, causing a loss of information. 
We propose a new method to generate fake X-vectors which overcomes these limitations by preserving the distributional properties of X-vectors and their intra-similarity.
We use population data to learn the properties of the X-vector space, before fitting a generative model which we use to sample fake X-vectors.
We show how this approach generates X-vectors that more closely follow the expected intra-similarity distribution of organic speaker X-vectors.
Our method can be easily integrated with others as the anonymization component of the system and removes the need to distribute a pool of speakers to use during the anonymization.
Our approach leads to an increase in EER of up to 19.4\% in males and 11.1\% in females in scenarios where enrollment and trial utterances are anonymized versus the baseline solution, demonstrating the diversity of our generated voices.

\end{abstract}

\section{Introduction}
Recent advances in voice cloning have lead to extremely realistic synthetic voices~\cite{shen2018natural, Liu2018} and have shown how few voice samples are actually required to bypass voice authentication systems~\cite{arik2018neural, turner2019attacking}.
As voice is personally identifiable, protecting voice data privacy from leaks and adversaries is necessary.

Speech anonymization is a novel technique which aims to anonymize voice data while retaining both the words spoken in the audio and the way they are spoken, such as tone and delivery.
Ideally, anonymized voices need to be different from the voice of the original speaker, in a way that guarantees that an anonymized voice is not linkable to the original speaker.
The VoicePrivacy Challenge 2020~\cite{VoicePriv2020}, which this paper is part of, aims to drive forward the creation of voice anonymization systems.
This is done by providing a common set of datasets, a baseline anonymization system and a framework for assessing the performance of voice anonymization.

In this paper we note that the  X-vector anonymization proposed in the VoicePrivacy Challenge 2020 baseline system leads to fake X-vectors which underutilize their multi-dimensional vector space.
Consequently, they tend to be very similar to each other, leading to anonymized voices that are similar as a result.
We show how this abnormal similarity is evident by looking at the distribution of cross-similarity between pairs of X-vectors, comparing the distribution of fake and original similarities.
We then propose a method that better leverages the vector space by learning the properties of this space from population data and fitting a generative model on a reduced-dimensionality space.
The generative model is then used to sample fake X-vectors for anonymization.
We show the performance of our method comparing it to the baseline, showing how our generation method significantly improves the diversity between anonymous voices while retaining the baseline performance across other metrics.
Our main contributions are as follows:
\begin{enumerate}[leftmargin=.55cm]
  \item We analyze the shortcomings of the baseline anonymous X-vector generation method, showing that generated X-vectors tend to be much more similar to each other than original X-vectors are.
  \item We present a general method that improves on the anonymous X-vector generation by learning the distributional properties of the X-vector space and by fitting a generative model on this space, where X-vectors can be sampled from. Our method also removes the requirement of having a pool of speakers' X-vectors during the anonymization process (as in the baseline), which may lead to privacy leakages.
  \item We evaluate our method within the VoicePrivacy Challenge 2020, showing its improved performance in creating differing anonymous voices.
\end{enumerate}

\section{Related Work}
\subsection{Speaker Anonymization}
Speaker privacy is not a new concept, with works on securing and encrypting voices existing for several decades, dating back to the analog processing era~\cite{Cox1987}.
This physical layer anonymization has its uses, but approaches that operate at this level either do not mask the voice itself, such as by adding a signal to existing audio~\cite{Hashimoto2016}, or render the audio unintelligble without a decryption key, preventing use for other legitimate purposes.

In this work we focus specifically on anonymization, in which personally identifiable attributes of the speech signal are supressed, but leaving intact all other aspects.
Past work in this area includes using voice transformation to convert voices to a specific special speaker identity~\cite{Jin2009} or using a Convolutional Neural Network (CNN) to convert each speaker to a new anonmyous voice, created as a function of a set of transformation features between the source voice and a database of voices~\cite{Bahmaninezhad2018}.
The level of anonymization offered by previous work is not immediately clear, and hence the VoicePrivacy Challenge 2020 has been created to evaluate systems with common datasets, protcols and metrics~\cite{VoicePrivacyInitiative}.

\subsection{The VoicePrivacy Challenge 2020}
The VoicePrivacy Challenge 2020~\cite{VoicePriv2020} provides the setting for this paper, and defines a specific goal, set of datasets, and set of metrics for the evaluation and comparison of voice anonymization systems.
The challenge seeks solutions for a scenario where `Speakers want to hide their identity whilst still allowing all other downstream goals to be achieved'~\cite{VoicePrivacyInitiative}. This is done by converting a speaker to a \textit{pseudo-speaker}, the new identity of the original speaker.

In order to meet the task of achieving downstream goals the following system requirements are given: (a) output a speech waveform, (b) hide speaker identity as much as possible, (c) distort other speech characteristics as little as possible, (d) ensure that all trial utterances from a given speaker appear to be uttered by the same pseudo-speaker, while trial utterances from different speakers appear to be uttered by different pseudo-speakers.

The challenge provides a common set of permitted datasets, to facilitate an even playing field. 
Likewise it provides the evaluation framework, consisting of a set of objective metrics presented in this work, as well as subjective metrics calculated by the challenge organisers in the future.

\subsection{Threat Model}
The challenge assumes that the attackers have access to one or more anonymized trial utterances, and possibly also to original or anonymized enrollment utterances for each speaker.
The threat model states that the attacker does not have access to the anonymization system applied by the user.
Whilst our submission operates under this threat model, we do not believe this assumption is necessarily the most reasonable for a speaker anonymization system.
In fact, within the security field it is typical to assume that an attacker knows the details of the system (Kerckhoffs's principle).
\section{System Overview}
\subsection{Rationale}\label{sec:rationale}
\begin{figure}[t]
	\centering
	\includegraphics[width=\columnwidth]{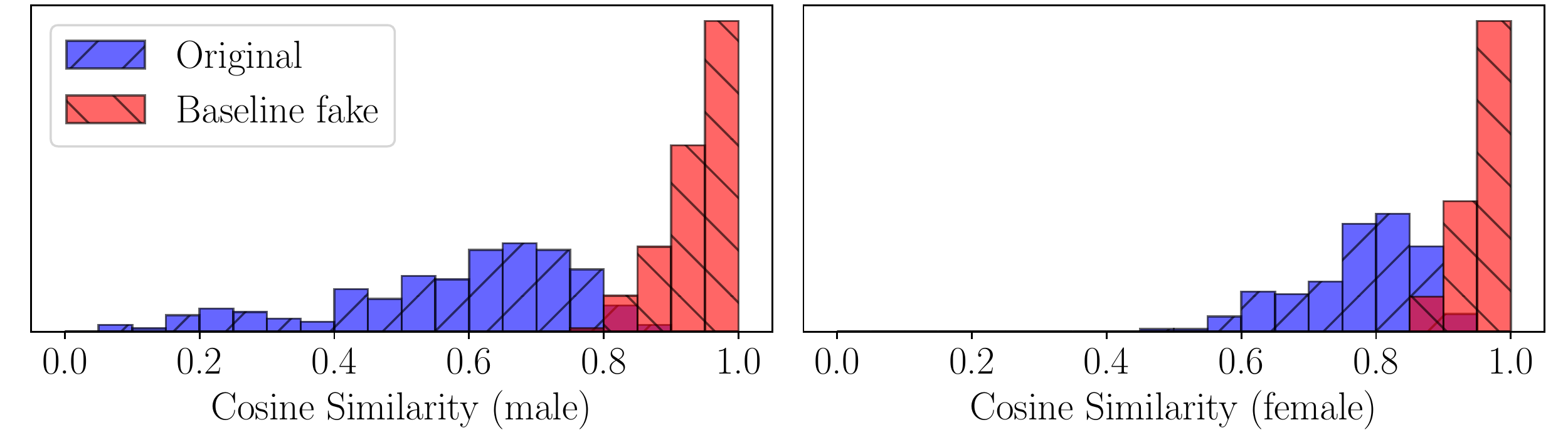}
	\caption{Distribution of cross-cosine similarities between pairs of X-vectors from original voices and from the baseline fakes. The baseline fake X-vectors do not follow the same distribution of cosine similarities as the original X-vectors: these fake X-vectors are much more similar to one another than X-vectors extracted from organic speakers.}
	\label{Fig:OriginalsVsFakes}
\end{figure}
Our system design follows the same approach as that of~\cite{Fang2019}.
Fang et al.~\cite{Fang2019} proposed three techniques for generating fake X-vectors: (i) nearest speakers, (ii) random selection and (iii) range selection.
The VoicePrivacy Challenge 2020 baseline system uses a variant of the last of these techniques, selecting the 200 furthest away X-vectors, and then averaging a random selection of 100 within these.

The rationale behind this work is that the fake X-vectors generation techniques mentioned above introduce a bias which leads to their distribution being different from that of the original X-vector's.
In particular, we notice that mantaining the X-vector cross-similarity properties of the distribution is desirable, i.e., the similarity between fakes should have the same behavior as the similarity between originals.
We show in Figure~\ref{Fig:OriginalsVsFakes} the cross-similarity between each pair of X-vectors in the original pool and the fake pool.
From Figure~\ref{Fig:OriginalsVsFakes}, it is evident that the similarities in the fake pool are higher compared to the original pool\footnote{We also note that female X-vectors follow a different distribution than male ones, probably an effect of the unbalanced training data}.
This behavior comes as a consequence of averaging many X-vectors in the fake generation phase and leads to the global X-vector space being underutilized, which brings the following disadvantages:
\begin{itemize}[topsep=3pt,noitemsep,leftmargin=.5cm]
	\item less entropy in fake X-vector space: anonymized voices are more similar to each other than pairs of original voices,
	\item reduced privacy, as it's easier to tell anonymized and original voices apart.
	\item the system requires a pool of X-vectors to sample from in the anonymization phase, which may lead to privacy leaks as this information needs to be shipped with the system.
\end{itemize}
In the following we explain how our system improves on the X-vector generation to maintain the desired similarity properties.

%
%

\subsection{Method}
\begin{figure}[t]
\includegraphics[width=\columnwidth]{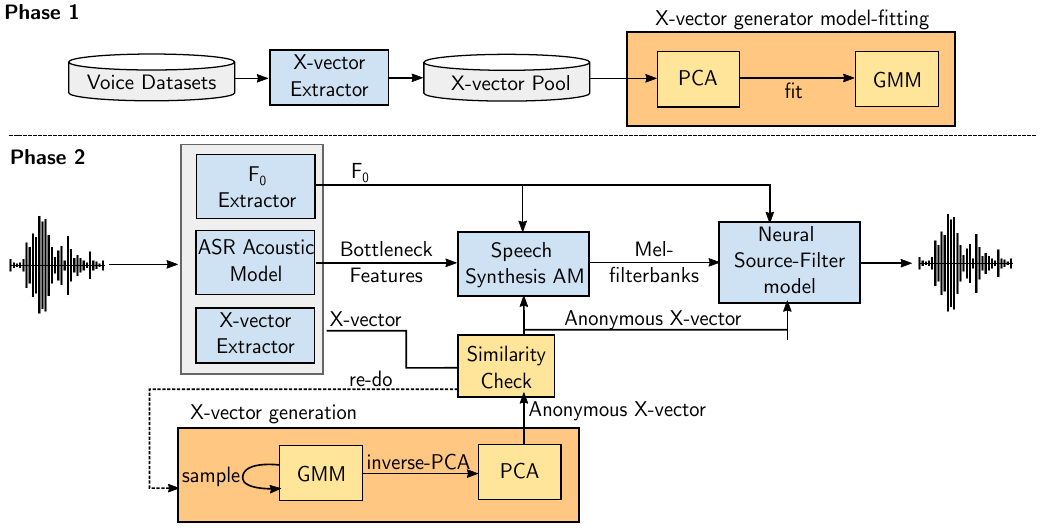}
\caption{Voice Anonymization system diagram. We replace the baseline X-vector anonymization module with a new generation method (shown in orange in the diagram).}
\label{Fig:SystemDiagram}
\end{figure}
In our method, we focus on improving the fake X-vector generation of the baseline system.
In the baseline, three types of features are extracted for a speaker, the fundamental frequency, bottleneck features and the X-vector.
The X-vector describes the speaker identity while the other features only encode the speech content~\cite{Fang2019}.

Following the weakness outlined in Section~\ref{sec:rationale}, we improve the X-vector generation in two steps.
At first, we learn the properties of the 512-dimensional X-vector space by using principal component analysis (PCA) on a large X-vector dataset.
Secondly, we fit a generative model on the PCA-reduced space, in order to sample from it, we use a Gaussian Mixture Model (GMM).
By using a generative model we avoid the bias introduced by the baseline fake X-vector generation, which generates them by averaging  subgroups of population vectors.
Whenever a voice needs to be anonymized, a reduced-dimensionality vector is randomly sampled from the GMM and then brought back into the 512-dimensional X-vector space by applying the PCA inverse transform.
We note that the X-vector generation could be addressed by training a generative adversarial network, however GMMs have been shown to better generalize the captured distributions~\cite{Richardson2018} and do not suffer from membership inference attacks (which could harm the system privacy guarantees)~\cite{shokri2017membership}.
In Section~\ref{sec:gmm_fit} we show how we choose parameters for our method by monitoring the cross-cosine similarity between pairs of fake and original X-vectors.

As in the baseline, in the later stages of the anonymization a Speech Synthesis acoustic model is used to generate Mel-filterbanks, which are fed with the F0 and new X-vector to
a Neural source-filter model to generate audio.
We train and reuse the models in the exact same way as the baseline for this, with the exception that we use the VoxCeleb1~\cite{nagrani2017voxceleb} and VoxCeleb2~\cite{chung2018voxceleb2} datasets in our pool of speaker X-vectors, in addition to the LibriTTS~\cite{zen2019libritts} train-other-500 dataset.
Figure \ref{Fig:SystemDiagram} gives a full overview of how these system components fit together.

\subsection{Forced Dissimilarity}
Occasionally, our method might generate an X-vector from the GMM which is relatively close to the original user's voice, which may comprimise their anonymity.
To mitigate this risk we propose an optional similarity check, termed \textit{forced dissimilarity}, between the speaker's X-vector and the newly generated fake X-vector.
If the cosine distance between the two X-vectors is above some threshold, then a new fake X-vector is generated.
This operation adds almost no overhead to the system as it only requires a repeated sample from the GMM.

\section{Experiments}
\begin{figure}[t]
	\centering
	\includegraphics[width=\columnwidth]{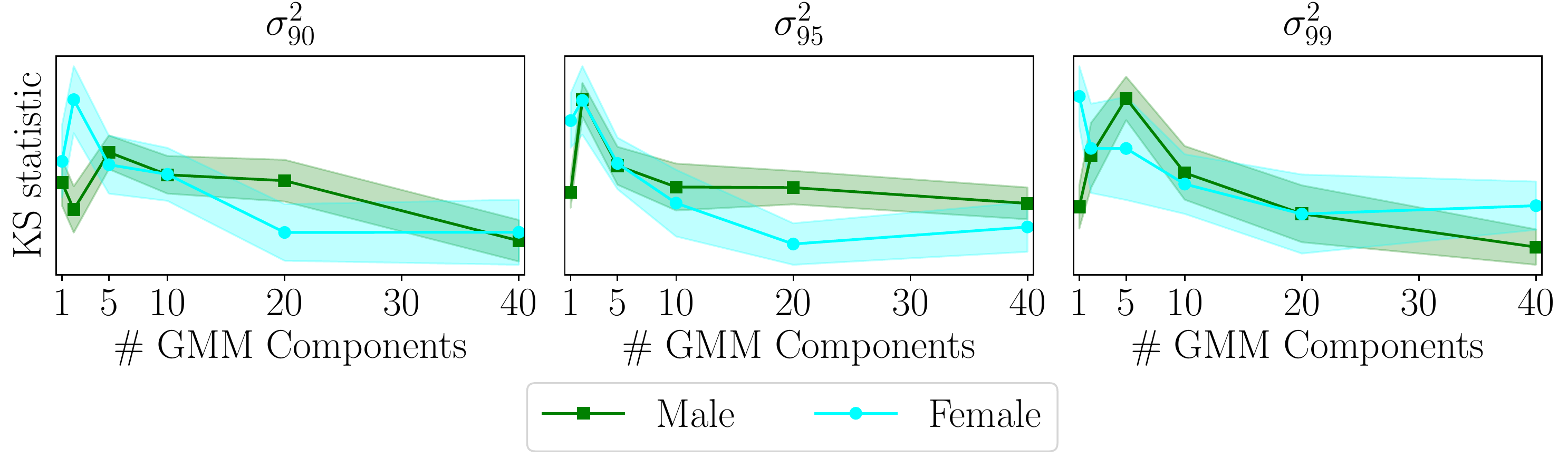}
	\caption{KS statistic between cross-similarity distributions of our fake X-vectors and VoxCeleb X-vectors, for varied PCA retained variance and GMM no. of components, per gender.}
	\label{fig:gmmFitVsVarDiag}
\end{figure}
\subsection{Determining Optimal Parameters}
\label{sec:gmm_fit}
\subsubsection{Setup}
To evaluate the performance of our fake X-vector generation, we perform an analysis on the resulting cross-similarity distribution of the generated vectors while varying the number of PCA and GMM components.
We monitor how close the cross-similarity distribution between the fake and original X-vectors are (as in Figure~\ref{Fig:OriginalsVsFakes}) using the Kolmogorov-Smirnov (KS) test between the two distributions.
The KS test quantifies the distance between two empirical cumulative distribution functions (eCDF), with lower scores implying that two distributions are more similar.
For PCA, we focus on three values for the total amount of variance captured, namely 90\%, 95\% and 99\%.

We setup the evaluation as follows, we extract all the X-vectors from VoxCeleb1, VoxCeleb2 (4,451 and 2,912 for male and female), for each gender we perform a 50\% train-test split and we train our PCA+GMM on the training part.
We then sample from the GMM and apply the PCA inverse transform to obtain 512-dimensional fake X-vectors.
Then we compute the distribution of cross-cosine similarity on fake X-vectors and on the remaining 50\% testing split, and we compute the KS statistic between the distributions.
For the GMM we learn a diagonal covariance matrix, set the maximum number of EM iterations to 1,000 and the convergence tolerance to $10^{-16}$.

\subsubsection{Results}

\begin{figure}[t]
	\centering
	\includegraphics[width=\columnwidth]{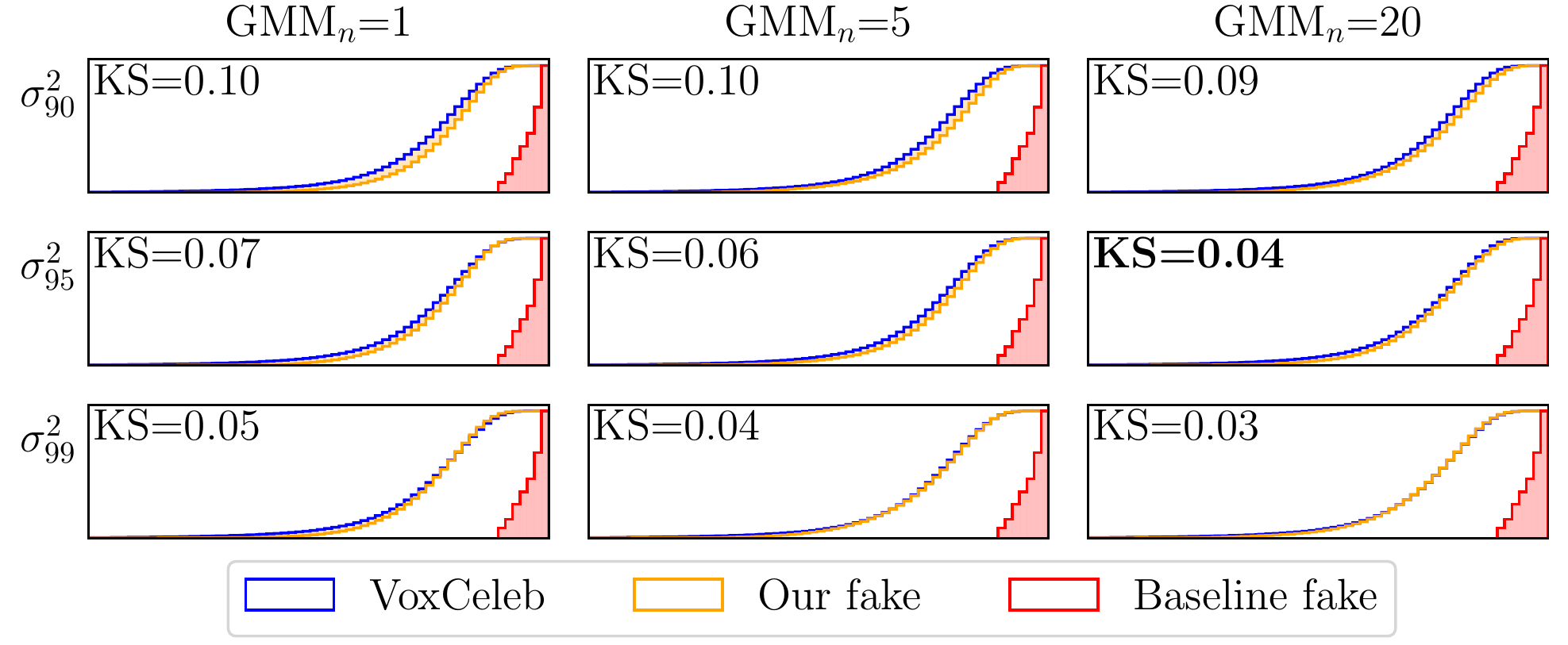}
	\caption{eCDF comparison between our fake, Voxceleb test split and baseline X-vectors, for female speakers (male distributions are similar). The combination of parameters chosen for the evaluation is in bold. KS values are between VoxCeleb and our fakes, and are only directly comparable across rows.}
	\label{fig:ecdfComparison}
\end{figure}

We report in Figure~\ref{fig:gmmFitVsVarDiag} the results for the three PCA models with increasing number of components used to fit the GMM, and we report some examples of the resulting eCDFs in Figure~\ref{fig:ecdfComparison}.
We found that using one or two GMM component(s) lead to a relatively good fit for males but not for females, where there is a clearer decreasing trend as the no. of GMM components increases.
Figure~\ref{fig:ecdfComparison} shows how much closer our fake X-vectors approximate the cross-similarity distributions found in the VoxCeleb data compared to the baseline fake X-vectors.
While increasing the number of components generally leads to a greater similarity between the distributions, in order to not overfit to the VoxCelebs data we choose to settle on using 95\% of PCA-retained variance and 20 GMM components.
This allows us to have a good approximation of the 512-dimensional X-vector space without requiring an overly complicated model.

\subsection{VoicePrivacy Challenge 2020 Results}

	\begin{table*}[t]
\centering
\footnotesize
\caption{Speaker verifiability results for the pretrained ASV\textsubscript{eval} model.
The table shows the results for our anonymization method with 20 GMM components and $\sigma^2_{95}$ PCA, without forced distancing.
In parenthesis we report the difference with the baseline system.
}
\begin{tabular}{c|c|c|c|c|c|c|c|c|c}
\hline
\multirow{2}{*}{\textbf{Dataset}} & \multirow{2}{*}{\textbf{Gender}} & \multicolumn{2}{c|}{\textbf{Anonymization}}              & \multicolumn{3}{c|}{\textbf{Development}}             & \multicolumn{3}{c}{\textbf{Test}}                    \\ \cline{3-10} 
                                  &                                  & \textbf{Enroll}           & \textbf{Trial}              & \textbf{EER (\%)} & \textbf{C$_{\text{llr}}^{\text{min}}$} & \textbf{C\textsubscript{llr}} & \textbf{EER (\%)} & \textbf{C$_{\text{llr}}^{\text{min}}$} & \textbf{C\textsubscript{llr}} \\ \hline\parbox[t]{2mm}{\multirow{6}{*}{\rotatebox[origin=c]{90}{LibriSpeech}}} & \multirow{3}{*}{Female} & \multirow{2}{*}{Original} & Original & 8.7 & 0.30 & 42.9 & 7.7 & 0.18 & 26.8 \\ \cline{4-10}
 &  &  & \multirow{2}{*}{Anonymized} & 44.2(-6.1) & 0.97(-0.03) & 139.8(-6.2) & 46.0(-2.5) & 0.99(-0.01) & 148.3(-3.1) \\ \cline{3-3} \cline{5-10}
 &  & Anonymized &  & 46.0(+10.9) & 0.98(+0.10) & 41.9(+26.7) & 40.9(+11.1) & 0.94(+0.14) & 26.0(+12.0) \\ \cline{2-10}
 & \multirow{3}{*}{Male} & \multirow{2}{*}{Original} & Original & 1.2 & 0.03 & 14.2 & 1.1 & 0.04 & 15.3 \\ \cline{4-10}
 &  &  & \multirow{2}{*}{Anonymized} & 48.9(-9.5) & 0.98(-0.02) & 154.8(-13.7) & 48.1(-5.1) & 0.97(-0.03) & 156.4(-10.7) \\ \cline{3-3} \cline{5-10}
 &  & Anonymized &  & 44.7(+15.1) & 0.95(+0.15) & 18.6(-1.5) & 41.9(+9.3) & 0.95(+0.11) & 21.9(-4.6) \\ \bottomrule\parbox[t]{2mm}{\multirow{6}{*}{\rotatebox[origin=c]{90}{VCTK (diff)}}} & \multirow{3}{*}{Female} & \multirow{2}{*}{Original} & Original & 2.9 & 0.10 & 1.1 & 4.9 & 0.17 & 1.5 \\ \cline{4-10}
 &  &  & \multirow{2}{*}{Anonymized} & 47.2(-2.9) & 0.96(-0.03) & 165.2(+2.3) & 47.3(-1.5) & 0.99(-0.01) & 149.6(+7.2) \\ \cline{3-3} \cline{5-10}
 &  & Anonymized &  & 30.8(+1.3) & 0.86(+0.05) & 16.1(+5.9) & 41.6(+7.4) & 0.97(+0.08) & 15.3(+3.0) \\ \cline{2-10}
 & \multirow{3}{*}{Male} & \multirow{2}{*}{Original} & Original & 1.4 & 0.05 & 1.2 & 2.1 & 0.07 & 1.8 \\ \cline{4-10}
 &  &  & \multirow{2}{*}{Anonymized} & 45.1(-10.3) & 0.98(-0.02) & 151.2(-15.3) & 48.3(-5.4) & 0.99(-0.01) & 160.6(-5.0) \\ \cline{3-3} \cline{5-10}
 &  & Anonymized &  & 45.5(+19.4) & 0.99(+0.23) & 19.2(+0.4) & 45.1(+19.2) & 0.96(+0.22) & 33.4(+17.1) \\ \bottomrule\end{tabular}

\label{tab:one}
\end{table*}

\begin{table*}[t]
\centering
\footnotesize
\caption{Speaker verifiability results for the pretrained ASV\textsubscript{eval} model.
The table shows the results for our anonymization method with 20 GMM components and $\sigma^2_{95}$ PCA, with FD at a threshold of 0.8.
In parenthesis we report the difference with the baseline system.
}
\begin{tabular}{c|c|c|c|c|c|c|c|c|c}
\hline
\multirow{2}{*}{\textbf{Dataset}} & \multirow{2}{*}{\textbf{Gender}} & \multicolumn{2}{c|}{\textbf{Anonymization}}              & \multicolumn{3}{c|}{\textbf{Development}}             & \multicolumn{3}{c}{\textbf{Test}}                    \\ \cline{3-10} 
                                  &                                  & \textbf{Enroll}           & \textbf{Trial}              & \textbf{EER (\%)} & \textbf{C$_{\text{llr}}^{\text{min}}$} & \textbf{C\textsubscript{llr}} & \textbf{EER (\%)} & \textbf{C$_{\text{llr}}^{\text{min}}$} & \textbf{C\textsubscript{llr}} \\ \hline\parbox[t]{2mm}{\multirow{6}{*}{\rotatebox[origin=c]{90}{LibriSpeech}}} & \multirow{3}{*}{Female} & \multirow{2}{*}{Original} & Original & 8.7 & 0.30 & 42.9 & 7.7 & 0.18 & 26.8 \\ \cline{4-10}
 &  &  & \multirow{2}{*}{Anonymized} & 43.8(-6.5) & 0.96(-0.04) & 138.8(-7.2) & 48.9(+0.4) & 0.99(-0.00) & 154.4(+3.0) \\ \cline{3-3} \cline{5-10}
 &  & Anonymized &  & 37.8(+2.7) & 0.92(+0.04) & 25.1(+9.9) & 46.2(+16.4) & 0.99(+0.19) & 44.2(+30.2) \\ \cline{2-10}
 & \multirow{3}{*}{Male} & \multirow{2}{*}{Original} & Original & 1.2 & 0.03 & 14.2 & 1.1 & 0.04 & 15.3 \\ \cline{4-10}
 &  &  & \multirow{2}{*}{Anonymized} & 51.2(-7.1) & 0.97(-0.03) & 150.7(-17.8) & 49.7(-3.6) & 0.97(-0.03) & 157.1(-10.0) \\ \cline{3-3} \cline{5-10}
 &  & Anonymized &  & 43.2(+13.5) & 0.95(+0.14) & 17.6(-2.5) & 44.8(+12.2) & 0.96(+0.13) & 27.1(+0.6) \\ \bottomrule\parbox[t]{2mm}{\multirow{6}{*}{\rotatebox[origin=c]{90}{VCTK (diff)}}} & \multirow{3}{*}{Female} & \multirow{2}{*}{Original} & Original & 2.9 & 0.10 & 1.1 & 4.9 & 0.17 & 1.5 \\ \cline{4-10}
 &  &  & \multirow{2}{*}{Anonymized} & 48.0(-2.0) & 0.97(-0.02) & 168.7(+5.8) & 45.1(-3.8) & 0.99(-0.01) & 147.2(+4.8) \\ \cline{3-3} \cline{5-10}
 &  & Anonymized &  & 39.1(+9.7) & 0.89(+0.08) & 17.9(+7.7) & 42.1(+7.9) & 0.96(+0.07) & 16.3(+4.0) \\ \cline{2-10}
 & \multirow{3}{*}{Male} & \multirow{2}{*}{Original} & Original & 1.4 & 0.05 & 1.2 & 2.1 & 0.07 & 1.8 \\ \cline{4-10}
 &  &  & \multirow{2}{*}{Anonymized} & 47.3(-8.0) & 0.99(-0.01) & 154.5(-12.0) & 48.8(-4.9) & 1.00(-0.00) & 158.4(-7.2) \\ \cline{3-3} \cline{5-10}
 &  & Anonymized &  & 38.7(+12.6) & 0.89(+0.14) & 14.7(-4.2) & 49.5(+23.6) & 0.96(+0.22) & 30.7(+14.4) \\ \bottomrule\end{tabular}

\label{tab:two}
\end{table*}
\begin{table}[t]
	\footnotesize
\caption{WER rates for original and anonymized voices, with and without forced dissimilarity at a threshold of 0.8. In parenthesis we report the difference with the baseline system.}
\begin{tabular}{c|c|c|c}
\hline
& & \multicolumn{2}{c}{\textbf{WER (\%)} }  \\ \hline
\textbf{Dataset}                    & \textbf{Anonymization}        & \textbf{Dev.} & \textbf{Test} \\ \hline
\multirow{3}{*}{LibriSpeech}        & Original                      &      3.83                  &         4.14               \\ \cline{2-4} 
                                        & Anonymized                    &         9.24 (+2.74)               &           12.68 (+5.91)             \\ \cline{2-4} 
                                    & Anonymized with FD &             11.89 (+5.39)           &               9.38 (+2.61)         \\ \hline
\multirow{3}{*}{VCTK} & Original                      &       10.79                 &         12.81               \\ \cline{2-4}
                                    & Anonymized                    &             17.31 (+1.81)           &               15.96 (+0.43)         \\ \cline{2-4} 
                                    & Anonymized with FD &             16.35 (+0.85)           &               16.65 (+1.12)         \\ \bottomrule
\end{tabular}
	\label{Table:WER}
\end{table}

We focus mainly on Equal Error Rates (EER) and $\text{C}_{\text{llr}}^{\text{min}}$ in our analysis, as $\text{C}_{\text{llr}}$ is more ambiguous due to non-calibration~\cite{VoicePrivacyInitiative}.
Table~\ref{tab:one} and Table~\ref{tab:two} show the evaluation results computed our method, with and without forced dissimilarity (FD), respectively\footnote{A previous version of this work had a bug leading to incorrect results (slight improvement for original vs anonymised, degredation for anonymised vs anonymised compared to these results).}.
Both tables also show the comparison to the baseline\footnote{We omit results for VCTK (common) due to space constraints}.
For FD, we set the similarity threshold to 0.8.

We find that our results for both versions of our method perform similarly.
For scenarios where original enrollment and anonymized trial data are evaluated we achieve EERs of 44.2-47.3\% for females and 45.1-48.9\% for males on our standard system.
These EER's and corresponding $\text{C}_{\text{llr}}^{\text{min}}$ are slightly lower than the baseline solution, but are still close to 50\%, which would indicate perfect anonymization.

In scenarios where both enrollment and trial data are anonymized we achieve EERs of 30.8-46.0\% for females and 41.9-45.5\% for males, for our standard system.
For males this represent an improvement over the baseline of 9.3-19.4\%, and 1.3-11.1\% for females.
$\text{C}_{\text{llr}}^{\text{min}}$ values follow a similar pattern of improvement.
We believe the disparity in improvement for males and females may be due to the improved performance of the X-vector extractor on males, as seen by the wider distribution of similarities in Figure \ref{Fig:OriginalsVsFakes}.

The improvements in EER and $\text{C}_{\text{llr}}^{\text{min}}$ for scenarios with enrollment and trials anonymized show that our technique improves the heterogeneity of the anonymized voices.
It is likely that some of the remaining difference between our results and a 50\% EER is due to the X-vector system not perfectly decomposing the voice into speaker identity and non-speaker identity components.
However, our method is general enough that it could be applied to other systems which improve the voice identity extraction part.
	
Comparing our standard and FD methods, we notice that the results are similar, with the FD method having slightly better results for most dataset with only trials anonymized, and mixed results in trial and enroll anonymized scenarios, generally improving for test datasets and worsening for development datasets.
We do not expect the FD method to significantly impact our results, as it is intend as a measure to prevent transformations to anonymized voices that are very similar to the original voice, as opposed to improving results.

We observe that the word error rates (WER) for both our standard and FD systems are in the same regions as the baseline, with a degredation in quality for both datasets, although a smaller degredation for the VCTK dataset (maximum of 1.81\%), as shown in Table~\ref{Table:WER}.

\section{Conclusions}
In this work, we propose a scheme to anonymize voices in a way that better maintains the natural diversity of voices, as compared to previous approaches.
We maintain this diversity by learning the properties of the X-vector space and using a generative model to sample from it, showing that such method better captures the distribution of similarities between fake vectors.
This increase in the diversity of anonymized voices makes them more distinguishable from one another, as evidenced by improved results in scenarios where both enrollment and trial are anonymized.
In our work we also propose to use forced dissimilarity, which allows a speaker to ensure that the anonymized voice they produce is not too similar to their own voice.

We experimentally validate that our proposed system produces voices that are more diverse, and evaluate our system against the VoicePrivacy Challenge 2020 baseline system.
Our results in the challenge show a slight degredation in performance of anonymized voices against the original enrolled voices, but show a strong improvement when comparing two versions of the same persons voice anonymized.
Our results also reveal worse performance for females than males, which we believe to be a result of the unbalanced dataset used for training, and highlight the opportunity to improve such bias.

\section{Acknowledgements}
This work was generously supported by a grant from Mastercard  and by the Engineering and Physical Sciences Research Council [grant numbers EP/N509711/1, EP/P00881X/1]

\bibliographystyle{IEEEtran}

\balance

\bibliography{mybib}

\begin{thebibliography}{10}
\providecommand{\url}[1]{#1}
\csname url@samestyle\endcsname
\providecommand{\newblock}{\relax}
\providecommand{\bibinfo}[2]{#2}
\providecommand{\BIBentrySTDinterwordspacing}{\spaceskip=0pt\relax}
\providecommand{\BIBentryALTinterwordstretchfactor}{4}
\providecommand{\BIBentryALTinterwordspacing}{\spaceskip=\fontdimen2\font plus
\BIBentryALTinterwordstretchfactor\fontdimen3\font minus
  \fontdimen4\font\relax}
\providecommand{\BIBforeignlanguage}[2]{{%
\expandafter\ifx\csname l@#1\endcsname\relax
\typeout{** WARNING: IEEEtran.bst: No hyphenation pattern has been}%
\typeout{** loaded for the language `#1'. Using the pattern for}%
\typeout{** the default language instead.}%
\else
\language=\csname l@#1\endcsname
\fi
#2}}
\providecommand{\BIBdecl}{\relax}
\BIBdecl

\bibitem{shen2018natural}
J.~Shen, R.~Pang, R.~J. Weiss, M.~Schuster, N.~Jaitly, Z.~Yang, Z.~Chen,
  Y.~Zhang, Y.~Wang, R.~Skerrv-Ryan \emph{et~al.}, ``Natural tts synthesis by
  conditioning wavenet on mel spectrogram predictions,'' in \emph{2018 IEEE
  International Conference on Acoustics, Speech and Signal Processing
  (ICASSP)}.\hskip 1em plus 0.5em minus 0.4em\relax IEEE, 2018, pp. 4779--4783.

\bibitem{Liu2018}
L.~J. Liu, Z.~H. Ling, Yuan-Jiang, Ming-Zhou, and L.~R. Dai, ``{Wavenet vocoder
  with limited training data for voice conversion},'' \emph{Proceedings of the
  Annual Conference of the International Speech Communication Association,
  INTERSPEECH}, vol. 2018-Septe, no. September, pp. 1983--1987, 2018.

\bibitem{arik2018neural}
S.~Arik, J.~Chen, K.~Peng, W.~Ping, and Y.~Zhou, ``Neural voice cloning with a
  few samples,'' in \emph{Advances in Neural Information Processing Systems},
  2018, pp. 10\,019--10\,029.

\bibitem{turner2019attacking}
H.~Turner, G.~Lovisotto, and I.~Martinovic, ``Attacking speaker recognition
  systems with phoneme morphing,'' in \emph{European Symposium on Research in
  Computer Security}.\hskip 1em plus 0.5em minus 0.4em\relax Springer, 2019,
  pp. 471--492.

\bibitem{VoicePriv2020}
\BIBentryALTinterwordspacing
N.~Tomashenko, B.~M.~L. Srivastava, X.~Wang, E.~Vincent, A.~Nautsch,
  J.~Yamagishi, N.~Evans, J.~Patino, J.-F. Bonastre, P.-G. No{\'e}, and
  M.~Todisco, ``The {VoicePrivacy} 2020 {Challenge} evaluation plan,'' 2020.
  [Online]. Available:
  \url{https://www.voiceprivacychallenge.org/docs/VoicePrivacy_2020_Eval_Plan_v1_3.pdf}
\BIBentrySTDinterwordspacing

\bibitem{Cox1987}
R.~V. Cox, D.~E. Bock, K.~B. Bauer, J.~D. Johnston, and J.~H. Synder, ``{Analog
  Voice Privacy System.}'' \emph{AT{\&}T Technical Journal}, vol.~66, no.~1,
  pp. 119--131, 1987.

\bibitem{Hashimoto2016}
K.~Hashimoto, J.~Yamagishi, and I.~Echizen, ``{Privacy-preserving sound to
  degrade automatic speaker verification performance},'' \emph{ICASSP, IEEE
  International Conference on Acoustics, Speech and Signal Processing -
  Proceedings}, vol. 2016-May, pp. 5500--5504, 2016.

\bibitem{Jin2009}
Q.~Jin, A.~R. Toth, T.~Schultz, and A.~W. Black, ``{Speaker de-identification
  via voice transformation},'' \emph{Proceedings of the 2009 IEEE Workshop on
  Automatic Speech Recognition and Understanding, ASRU 2009}, pp. 529--533,
  2009.

\bibitem{Bahmaninezhad2018}
F.~Bahmaninezhad, C.~Zhang, and J.~Hansen, ``{Convolutional Neural Network
  Based Speaker De-Identification},'' vol. 2016, no. June, pp. 255--260, 2018.

\bibitem{VoicePrivacyInitiative}
N.~Tomashenko, B.~M.~L. Srivastava, X.~Wang, E.~Vincent, A.~Nautsch,
  J.~Yamagishi, N.~Evans, J.~Patino, J.-F. Bonastre, P.-G. No{\'e}, and
  M.~Todisco, ``Introducing the {VoicePrivacy} initiative,'' 2020.

\bibitem{Fang2019}
\BIBentryALTinterwordspacing
F.~Fang, X.~Wang, J.~Yamagishi, I.~Echizen, M.~Todisco, N.~Evans, and J.-F.
  Bonastre, ``{Speaker Anonymization Using X-vector and Neural Waveform
  Models},'' pp. 3--8, 2019. [Online]. Available:
  \url{http://arxiv.org/abs/1905.13561}
\BIBentrySTDinterwordspacing

\bibitem{Richardson2018}
E.~Richardson and Y.~Weiss, ``{On GANs and GMMs},'' in \emph{Advances in Neural
  Information Processing Systems}, no. NeurIPS, 2018, pp. 5847--5858.

\bibitem{shokri2017membership}
R.~Shokri, M.~Stronati, C.~Song, and V.~Shmatikov, ``Membership inference
  attacks against machine learning models,'' in \emph{2017 IEEE Symposium on
  Security and Privacy (SP)}.\hskip 1em plus 0.5em minus 0.4em\relax IEEE,
  2017, pp. 3--18.

\bibitem{nagrani2017voxceleb}
\BIBentryALTinterwordspacing
A.~Nagrani, J.~S. Chung, and A.~Zisserman, ``Voxceleb: A large-scale speaker
  identification dataset,'' in \emph{Proc. Interspeech 2017}, 2017, pp.
  2616--2620. [Online]. Available:
  \url{http://dx.doi.org/10.21437/Interspeech.2017-950}
\BIBentrySTDinterwordspacing

\bibitem{chung2018voxceleb2}
\BIBentryALTinterwordspacing
J.~S. Chung, A.~Nagrani, and A.~Zisserman, ``Voxceleb2: Deep speaker
  recognition,'' in \emph{Proc. Interspeech 2018}, 2018, pp. 1086--1090.
  [Online]. Available: \url{http://dx.doi.org/10.21437/Interspeech.2018-1929}
\BIBentrySTDinterwordspacing

\bibitem{zen2019libritts}
\BIBentryALTinterwordspacing
H.~Zen, V.~Dang, R.~Clark, Y.~Zhang, R.~J. Weiss, Y.~Jia, Z.~Chen, and Y.~Wu,
  ``{LibriTTS: A Corpus Derived from LibriSpeech for Text-to-Speech},'' in
  \emph{Proc. Interspeech 2019}, 2019, pp. 1526--1530. [Online]. Available:
  \url{http://dx.doi.org/10.21437/Interspeech.2019-2441}
\BIBentrySTDinterwordspacing

\end{thebibliography}


\end{document}